\documentclass[twocolumn]{aastex701}

\usepackage[utf8]{inputenc}

\usepackage{color}
\usepackage{siunitx}
\newcommand{\rev}[1]{#1}
\begin{document}

\title{The START mission: Probing the 2029 Apophis Flyby via a Low-Cost Geocentric SmallSat}

\author[orcid=0009-0004-6339-282X]{Yixuan Wu}
\altaffiliation{These authors contributed equally to this work as co-first authors.}
\affiliation{Tsinghua University, Beijing 100084, China}
\email[]{}

\author[orcid=0009-0007-2916-4470]{Yunda Li}
\altaffiliation{These authors contributed equally to this work as co-first authors.}
\affiliation{Tsinghua University, Beijing 100084, China}
\email[]{}

\author[]{Bowen Zheng}
\affiliation{Tsinghua University, Beijing 100084, China}
\email[]{}

\author[]{Jingxiang Xu}
\affiliation{Tsinghua University, Beijing 100084, China}
\email[]{}

\author[]{Yuhang Yu}
\affiliation{Tsinghua University, Beijing 100084, China}
\email[]{}

\author[]{Wang Sun}
\affiliation{Tsinghua University, Beijing 100084, China}
\email[]{}

\author[]{Haozhe Du}
\affiliation{Tsinghua University, Beijing 100084, China}
\email[]{}

\author[]{Siyuan Wu}
\affiliation{Tsinghua University, Beijing 100084, China}
\email[]{}

\author[]{Zhixuan Zhao}
\affiliation{Tsinghua University, Beijing 100084, China}
\email[]{}

\author[]{Rongchenzhong Hu}
\affiliation{Tsinghua University, Beijing 100084, China}
\email[]{}

\author[]{Jiantao Liu}
\affiliation{Tsinghua University, Beijing 100084, China}
\email[]{}

\author[]{Haoyu Liu}
\affiliation{Tsinghua University, Beijing 100084, China}
\email[]{}

\author[]{Yunchong Shen}
\affiliation{Tsinghua University, Beijing 100084, China}
\email[]{}

\author[]{Yutong Sun}
\affiliation{Tsinghua University, Beijing 100084, China}
\email[]{}

\author[]{Tianchen Wang}
\affiliation{Tsinghua University, Beijing 100084, China}
\email[]{}

\author[]{Jing Yu}
\affiliation{Tsinghua University, Beijing 100084, China}
\email[]{}

\author[]{Zijie Tang}
\affiliation{Tsinghua University, Beijing 100084, China}
\email[]{}

\author[]{Yihong Wu}
\affiliation{Tsinghua University, Beijing 100084, China}
\email[]{}

% \author[]{XXXs}
% \affiliation{Tsinghua University, Beijing 100084, China}
% \email[]{}

\author[orcid=0000-0003-1097-0521]{Yifei Jiao}
\affiliation{Tsinghua University, Beijing 100084, China}
\affiliation{University of California, Santa Cruz, CA 95064, USA}
\email[]{}

\author[orcid=0000-0002-0209-7286]{Wen-Yue Dai}
\affiliation{Tsinghua University, Beijing 100084, China}
\email[]{}

\author[orcid=0009-0003-7787-7847]{Jialong Song}
\affiliation{Tsinghua University, Beijing 100084, China}
\email[]{}

\author[orcid=0000-0002-9782-9282]{Nan Zhang}
\affiliation{Tsinghua University, Beijing 100084, China}
\affiliation{Beihang University, Beijing 100191, China}
\email[]{}

\author[orcid=0000-0003-2568-7948]{Zhaokui Wang}
\affiliation{Tsinghua University, Beijing 100084, China}
\email[]{}

\author[]{Tianshu Wang}
\affiliation{Tsinghua University, Beijing 100084, China}
\email[]{}

\author[]{Junfeng Li}
\affiliation{Tsinghua University, Beijing 100084, China}
\email[]{}

\author[orcid=0000-0002-6389-8677]{Hexi Baoyin}
\affiliation{Tsinghua University, Beijing 100084, China}
\affiliation{Inner Mongolia University of Technology, Hohhot 010051, China}
\email[show]{baoyin@tsinghua.edu.cn}

\author[orcid=0000-0002-8025-9113]{Bin Cheng}
\affiliation{Tsinghua University, Beijing 100084, China}
\email[show]{bincheng@tsinghua.edu.cn}

\begin{abstract}

\rev{The Student-led Threatening Asteroid Reconnaissance of Tsinghua (START) marks the inaugural mission of the Tianxing Program, a deep-space exploration initiative led by students at Tsinghua University. Conceived as a low-cost, rapid-response small-satellite mission, START will investigate the near-Earth asteroid (99942) Apophis during its close approach to Earth in April 2029, when significant terrestrial tidal forces will offer a rare opportunity to observe the physical and dynamical response of a rubble-pile asteroid in real time.} The \rev{\SI{300}{\kg}} spacecraft \rev{is scheduled for rideshare launch into low Earth orbit in early 2028 and will then use solar electric propulsion to transfer to} \rev{a circular mission orbit with a semimajor axis of approximately \SI{38000}{\km} and an inclination of \SI{55}{\degree}}. START will execute a high-speed flyby of Apophis from a closest distance of \SI{7}{\km} at the asteroid's closest approach to Earth, to capture its tidal response. The payload suite consists of a high-resolution narrow-field camera, a wide-field navigation camera, and \rev{three near-infrared multispectral} imagers. By acquiring centimeter-scale imagery and \rev{meter-scale} reflectance spectra, the mission will characterize the asteroid’s surface morphology and composition and \rev{monitor immediate tidally induced surface modifications}. Ultimately, START will provide geometrically complementary, high-cadence datasets to international flagship missions to Apophis, demonstrating a highly scalable and cost-effective architecture for future planetary defense and small-body exploration.

\end{abstract}



\section{Introduction}

The near-Earth asteroid (99942) Apophis is currently the focus of unprecedented international scrutiny owing to its exceptionally rare Earth close approach (ECA) on April 13, 2029 \citep{giorgini2008predicting}. 
During this event, the $\sim$\SI{340}{\m} diameter stony asteroid will pass within \SI{31600}{\km} of the Earth's surface—a distance closer than geostationary orbit. 
Such a deep penetration into Earth’s gravitational tidal field by an object of this size is a once-in-a-millennium event \citep{Binzel2021Apophis}, making Apophis visible to the naked eye for approximately two billion people in parts of Europe, Africa and Western Asia. 
This historic proximity serves as a powerful catalyst for global public engagement and has led the United Nations to declare 2029 the International Year of Asteroid Awareness and Planetary Defence \citep{michel2024international}. 
For the scientific community, this encounter represents more than a standard reconnaissance opportunity; while previous small-body missions have typically characterized asteroids as static, evolved remnants of the early solar system, the 2029 ECA provides the first-ever opportunity to witness active planetary evolution in real time. 
By observing this natural experiment, we can transition from analyzing the fossilized results of past processes to directly monitoring the immediate physical and dynamical responses of a rubble-pile body to terrestrial tidal forces. 
The anticipated tidal torques are expected to induce measurable changes in Apophis’s spin state and orientation, potentially triggering localized surface resurfacing, mass wasting, and seismic activity \citep{demartini2019using,yu2014numerical,kim2023tidal,ballouz2024shaking}. 
Capturing these ongoing transitions will yield an unprecedented scientific output, offering deep insights into \rev{the geotechnical properties of potentially hazardous asteroids and clues to their internal structure}.

Recognizing this opportunity, the international space community has mobilized a fleet of missions to characterize Apophis across different temporal and spatial scales. 
The European Space Agency (ESA) is developing the Rapid Apophis Mission for Space Safety (RAMSES), which aims to rendezvous with the asteroid in early 2029—approximately two months before ECA—to conduct a comprehensive ``before-and-after" survey \citep{kueppers2023ramses}. 
Following the encounter, NASA’s OSIRIS-APEX will rendezvous with Apophis to investigate its post-encounter state and conduct active surface-disturbance experiments \citep{dellagiustina2023osiris}. 
Furthermore, JAXA’s DESTINY+ mission, en route to 3200 Phaethon, is planned to perform a high-speed flyby of Apophis in early 2029, potentially providing the first close-up images to aid the subsequent rendezvous operations \citep{arai2025planetary}. 
Additional reconnaissance efforts are being explored through NASA's Request for Information to repurpose the Janus spacecraft for a pre-encounter flyby \citep{scheeres2024potential,scheeres2024janus,binzel2025findings}, as well as a growing number of innovative SmallSat concepts such as MIT Project Apophis \citep{binzel2017project,MITProjectApophisSET2017}, Caltech's mission concept \citep{daca2025caltech}, Argotec's ATENA \citep{saita2024apophis}, Milo Institute's Apophis Pathfinder \citep{bell2025apophis}, and Apophis Recon Swarm \citep{yu2025apophis}. 

While these flagship missions provide comprehensive global mapping, the scientific community would benefit significantly from multi-point, geometrically complementary datasets, particularly during the high-dynamic window around the ECA. 
\rev{Through its dedicated high-speed flyby, START will observe Apophis from an additional viewing geometry when terrestrial tidal forcing is strongest.} 

Beyond the scientific yield, the 2029 ECA introduces a rarely available engineering advantage: Apophis becomes temporarily accessible from geocentric trajectories. 
Unlike traditional asteroid missions that require multi-year deep-space cruise phases and \rev{high departure energies}, Apophis’s trajectory carries it deep into Earth’s gravity well. 
This provides a unique window for a minimalist flyby architecture, where a spacecraft can be deployed from Earth-orbiting platforms—such as low Earth orbit (LEO) or geostationary transfer orbit (GTO)—and reach the target within months using modest propulsion systems. 
This ``Earth-vicinity" accessibility effectively lowers the barriers to deep-space exploration, enabling rapid-response mission designs that were previously cost-prohibitive.

\rev{Leveraging this window, we propose the Student-led Threatening Asteroid Reconnaissance of Tsinghua (START), a low-cost, rapid-response flyby mission. As the inaugural mission of the Tianxing Program, it is being developed by Tsinghua University in collaboration with the Deep Space Exploration Laboratory and commercial space partners.}
START is designed to demonstrate a scalable and repeatable model for planetary defense and space exploration. 
Operating within a budget of approximately \rev{30} million RMB ($\sim$ \rev{$4.5$} million USD) and supported by private investment, the mission utilizes a \rev{\SI{300}{\kg}} small satellite platform launched as a secondary payload into low Earth orbit in early 2028. 
\rev{START will use solar electric propulsion (SEP) to reach its mission orbit and then perform a high-speed flyby of Apophis near ECA to investigate surface responses under strong terrestrial tidal forcing}. 
\rev{The nominal distance at START--Apophis closest approach (CA) is \SI{7}{\km}, balancing imaging performance with operational separation from RAMSES spacecraft (Section~\ref{4.1}).} 
Equipped with a high-resolution narrow-field camera (NFOV), a wide-field navigation camera (WFOV), and \rev{three near-infrared multispectral imagers (MSIs)} covering the \rev{\SIrange{0.7}{1.65}{\micro\m}} range, START will provide centimeter-scale imagery and \rev{meter-scale} reflectance spectra.
\rev{To achieve high-resolution staring imaging, these payloads are mounted on a two-axis, high-speed tracking mechanism featuring wide-range pointing and closed-loop tracking capabilities. Guided by an integrated data processor, the system utilizes real-time measurement data from the instruments to compute relative line-of-sight kinematics. This enables the tracking mechanism to stably track Apophis during the close-proximity flyby, effectively compensating for the extreme relative velocity, angular velocity, and angular acceleration.} 
These high-cadence observations are specifically designed to characterize the asteroid’s immediate physical and dynamical transitions—such as localized mass flow and \rev{spectral change}—providing critical constraints on \rev{regolith cohesion} and evolutionary pathways of rubble-pile near-Earth asteroids \rev{while complementing international flagship missions through a distinct observing geometry}.

This paper provides an overview of the START mission design and its current developmental status. 
The remainder of the article is structured as follows. 
Section~2 presents a high-level overview of the mission architecture. 
Section~3 details the primary scientific objectives and the supporting instrument suite. 
Section~4 describes the trajectory design, with a particular focus on the SEP-enabled transfer and precision flyby navigation strategies. 
Section~5 provides a technical description of the spacecraft platform and its subsystems.

\section{Mission overview}
%Jialong Song and Nan Zhang

\begin{figure*}[ht!]
    \centering
    \includegraphics[width=0.8\linewidth]{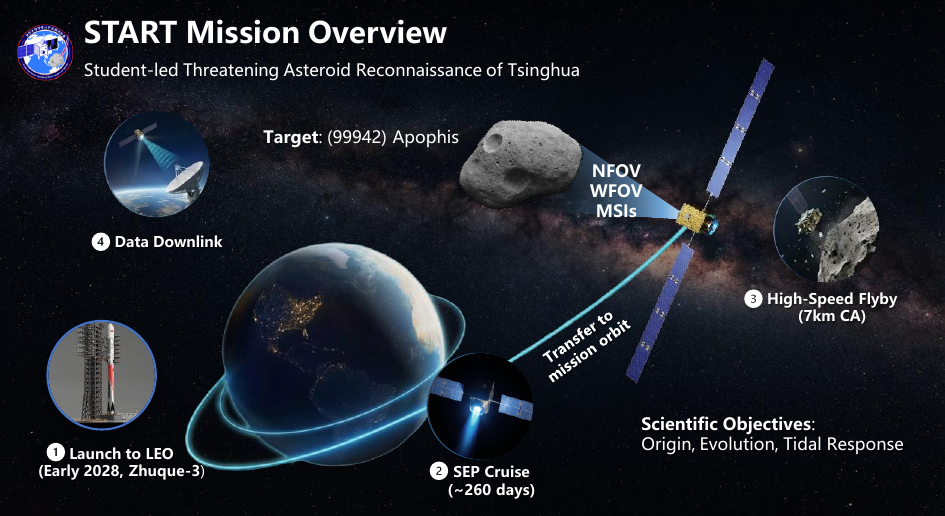}
    \caption{Overview of the START Mission}
    \label{fig:mission_overview}
\end{figure*}

START is designed to fly as a rideshare payload on a commercial launch vehicle and extensively utilizes commercial off-the-shelf (COTS) components to minimize cost. The nominal launch is scheduled for early 2028 aboard a LandSpace Zhuque-3 rocket. The spacecraft will initially be deployed into a \rev{\SI{500}{\km}-altitude LEO}. Subsequently, \rev{the spacecraft will transfer} to a circular mission orbit with \rev{a semimajor axis of approximately \SI{38000}{\km}} using its onboard \SI{60}{\mN} SEP system, matching \rev{Apophis's geocentric distance at ECA}. START is expected to reach its designated orbit \rev{at least one month} before the 2029 Apophis encounter \rev{for final checkout and rehearsals of target acquisition and tracking}. After executing trajectory correction maneuvers, START will perform a high-speed flyby of Apophis at a relative velocity of about \SI[per-mode = symbol]{8.74}{\km\per\s}. \rev{The observation window will begin $\sim$2~hours before Apophis's ECA as START approaches Apophis for its nominal \SI{7}{\km} CA, enabling observations of the asteroid's tidal response.}
\rev{Accounting for RAAN phasing in the parking orbit and the contingency margin during orbit raising, the nominal launch window lies in the first quarter of 2028.} Furthermore, any commercial launch vehicle with sufficient payload capacity can serve as a backup.
Figure~\ref{fig:mission_overview} presents a mission overview, with the operational phases defined as follows:

\begin{enumerate}
    \item \rev{\textbf{Launch, Commissioning, and RAAN Phasing (7--82 days):} Nominally scheduled for the first quarter of 2028, the spacecraft will be injected into a \SI{500}{\km}-altitude circular parking orbit. During the approximately 7-day commissioning phase, attitude acquisition and spacecraft checkout are conducted first, followed by initial in-orbit verification of payload functionality and performance. The spacecraft then remains in LEO as needed until $J_2$-driven nodal precession brings the RAAN to the required value.}
    \item \rev{\textbf{Orbit Transfer and Mission-Orbit Preparation (minimum transfer time of approximately 262 days):} Starting from the \SI{500}{\km} circular orbit, the \SI{60}{\mN} xenon SEP system raises the spacecraft to the target high Earth orbit. For the nominal \SI{300}{\kg} spacecraft, the transfer consumes approximately \SI{71.1}{\kg} of xenon and provides an equivalent velocity increment of approximately \SI[per-mode=symbol]{4.38}{\km\per\s}. The transfer is completed at least one month before CA, allowing final spacecraft checkout and rehearsals of target acquisition and tracking in the mission orbit.}
    \item \textbf{Detailed Observation ($\sim$2 hours):} Two hours prior to CA, \rev{Apophis will be initially acquired using the WFOV camera, enabling subsequent imaging by the high-resolution NFOV camera. The two-axis tracking mechanism will execute rapid pointing maneuvers to track the asteroid,} ensuring clear and stable observations by both the NFOV and the \rev{MSIs}. \rev{The NFOV high-resolution imaging window spans approximately \SI{57}{\s} to \SI{4}{\s} before CA, with the finest usable resolution reaching \SI{6.6}{\cm}/px.} The MSIs continue imaging closer to CA and achieve a best resolution of about \SI{1.9}{\m}/px. Thereafter, Apophis will gradually exit the NFOV, \rev{MSIs} and WFOV due to the \rev{tracking limitations of the tracking mechanism}.
    \item \textbf{Data Downlink Phase ($\sim$7 days):} Following the flyby observations, the spacecraft transmits the acquired scientific data to Earth. Data transmission initiates only after the spacecraft enters the communication window of the tracking stations.
\end{enumerate}

The START mission has completed its system-level design and transitioned into the detailed design phase. Finalization of the spacecraft's design is scheduled for late 2026, directly preceding the manufacturing phase. To date, the project has secured over \rev{20} million RMB in commercial funding and established preliminary technical partnerships with multiple commercial aerospace enterprises. Collectively, these milestones establish a robust foundation for the mission's subsequent developmental phases.

\section{Science objectives and instruments} 

The unique opportunity presented by Apophis's close encounter with Earth enables us to address two key science goals within a single mission: (1) What is the origin of Apophis and how did it evolve since its formation? (2) How does a close encounter with Earth modify the physical and dynamical states of Apophis? This section will provide a brief summary of the state-of-the-art understanding of these two science questions, and present our observation strategies and instruments to address them.

\rev{Within the broader international effort, START will observe Apophis near ECA while RAMSES accompanies the asteroid through the Earth encounter, supplementing observations from DESTINY+'s flyby (in early 2029) and OSIRIS-APEX's post-encounter survey. START's typical Sun--Apophis--observer phase angle of $\sim$\SI{10}{\degree} (compared with $\sim$\SI{20}{\degree} for RAMSES) and its distinct flyby azimuth together yield a viewing geometry complementary to that of RAMSES. START's optically focused payload is designed to maximize the spatial resolution of near-ECA imaging, with the NFOV camera reaching a peak resolution of approximately \SI{7}{\cm}/px, compared with RAMSES's \SI{10}{\cm} global-imaging objective \citep{kueppers2023ramses}. START's centimeter-scale images can then be compared with RAMSES imagery obtained before and after the encounter and with later OSIRIS-APEX observations, helping characterize encounter-induced surface changes. Subject to final geometry and coordination with the RAMSES team, START will seek opportunities to image Apophis together with the RAMSES spacecraft, creating a visual record of international cooperation in the exploration of Apophis.}

\subsection{Origin and evolution of Apophis}

Apophis is an Aten-type near-Earth asteroid with a semimajor axis slightly less than 1 au and a low inclination of \SI{3.3}{\degree} \citep{JPLapophis}. Its dynamical evolution is consistent with an origin in the inner main asteroid belt, followed by inward transport into near-Earth space likely driven by the Yarkovsky effect and the $\nu_6$ secular resonance \citep{morbidelli2003Yarkovsky,william2006Yarkovsky,reddy2018surface}. The high eccentricity of Apophis (e $\approx$ 0.19) places it near the upper end of the typical range for Aten objects, reflecting the dynamical excitation associated with its main-belt origin \citep{JPLapophis,bottke2002debiased,granvik2018debiased}.

Radar observations suggest that Apophis is an elongated, asymmetric, and possibly bifurcated asteroid, with an effective diameter of about \SI{340}{\m} consistent with both radar and thermal infrared estimates \citep{brozovic2018goldstone,satpathy2022NEOWISE}. It is spectrally classified as an Sq- or Q-type asteroid based on visible and near-infrared spectroscopy \citep{binzel2009spectral}, indicating a silicate-rich, ordinary-chondrite-like composition with limited space-weathering, characteristic of near-Earth asteroids and distinct from the more weathered S-type population dominant in the main belt \citep{vernazza2008compositional}. 

Apophis’s thermal inertia and its similarities to (25143) Itokawa in size, albedo, and composition suggest a rubble-pile internal structure rather than a monolithic body \citep{Muller2014thermal}. Such a structure is consistent with formation through collisional disruption of a larger parent body in the main belt, followed by gravitational re-accumulation of fragments \citep{Michel2001collisions,Kevin2018rubble}. In this scenario, Apophis represents a reassembled aggregate with significant macro-porosity, internal heterogeneity, and weak cohesive strength \citep{bottke2015asteroid}, making it particularly sensitive to external perturbations, such as tidal forces during close planetary encounters.

The rotational state of Apophis is complex, with photometric observations indicating a non-principal-axis or near-tumbling rotation \citep{pravec2014tumbling}, although its precise spin configuration remains uncertain. With a diameter of $\sim$\SI{340}{\m} and a rotation period of several tens of hours, Apophis lies in a regime where small asteroids are particularly susceptible to long-term modification of their spin states by the Yarkovsky–O’Keefe–Radzievskii–Paddack (YORP) effect \citep{william2006Yarkovsky}.

\subsection{Tide response of Apophis during Earth encounter}

The 2029 Earth close encounter of Apophis, with a geocentric distance of $\sim5.9R_{\rm E}$ at ECA, is not expected to cause catastrophic disruption or global resurfacing of the object \citep{yu2014numerical}. For reference, the fluid Roche limit is $\sim3.4R_{\rm E}$ assuming a typical bulk density of \SI{2000}{\kg\per\m\cubed} for S-type rubble-pile asteroids. The tidal forces, with a maximal magnitude of $\sim$\SI{1e-6}{\m\per\s\squared} during the encounter, are only about 1\% of the local surface gravity \citep{benson2023spin}.
On global scales, the tide forces could lead to an axis length change of about \SI{0.1}{\mm} compared with the body's dimension of \SI{340}{\m}, assuming elastic deformation under Earth's tidal forces \citep{demartini2019using}.

Nevertheless, the encounter would significantly change the orbit and spin state of Apophis. During the approach, Earth will perturb Apophis's orbit and change its semimajor axis from 0.92\,AU to 1.1\,AU \citep{dellagiustina2023osiris}, i.e., the post-encounter orbit will be an Apollo-class orbit.
The spin change of Apophis, however, is sensitive to the body's orientation at its ECA, which is difficult to predict at present. Current models suggest that its post-encounter rotation period may depart markedly from the pre-encounter value of \SI{30.56}{\hour}, and recent soft-sphere discrete element method simulations further place it at $\sim$\SIrange{18.6}{40.0}{\hour}, depending on the encounter geometry and internal structure \citep{demartini2019using,benson2023spin,ballouz2024shaking}. The pole direction of Apophis may likewise be reoriented by \SI{10}{\degree} or more, while deformation may introduce additional post-encounter attitude offsets of a few degrees relative to rigid-body predictions \citep{benson2023spin,hirabayashi2025prediction}.

Another physical outcome of great interest is the surface modification of Apophis during the encounter. Though global-scale resurfacing has been ruled out, we can still expect some local landslides on the scale of about 1\% of the entire surface area \citep{yu2014numerical,kim2023tidal}. The most likely resurfacing regions are the high slope regions ($\gtrsim$\SI{30}{\degree}), and the mid-slope regions with significant slope increases of up to $\sim$\SI{1}{\degree} \citep{kim2023tidal,valvano2022apophis}. \rev{The ``neck'' region of a bilobed body is one such area \citep{hirabayashi2021finite} and may be further perturbed by an internal mass shift \citep{wang2025simulating}.} \rev{Numerical simulations further indicate that the final hour before perigee offers the best opportunity to detect active tidal resurfacing \citep{kim2023tidal}.}
The local landslide may further cause boulder movements, regolith re-distribution, and also expose the fresh subsurface materials that remain unaltered by space weathering.
As a Sq-type asteroid, Apophis's Earth encounter will be an ideal natural experiment to help understand the correlations between S-, Sq-, Q-type asteroids and ordinary chondrite meteorites \citep{binzel2010earth}.
The encounter may also induce some discrete seismic events that lead to high-frequency ($>$\SI{0.1}{\Hz}) surface accelerations with magnitudes similar to Apophis's gravity \citep{ballouz2024shaking,demartini2025influence}.

\subsection{Science objectives and instruments}

As a low-cost, rapid-response mission, START is designed to characterize Apophis only during its close encounter with Earth in 2029.
On a geocentric trajectory, the spacecraft will perform a high-speed flyby of Apophis at a closest-approach distance of \SI{7}{\km}, but with only a short observation window ($\sim$\SI{2}{\hour}).
This is even more challenging than rendezvous missions.
Therefore, our mission will not directly measure the axial length change that requires a very high spatial resolution of \SI{0.1}{\mm} \citep{demartini2019using}, or the seismic activity that requires a landing seismometer.
\rev{Within these limitations, the science objectives and associated measurement requirements are summarized below.}

\begin{deluxetable*}{lllll}[htb!]
\tablecaption{Science Traceability Matrix for the START mission\label{tab:sci_obj}}
\tablewidth{0pt} % use natural width
\tabletypesize{\scriptsize}
\tablehead{
\colhead{Science \rev{questions}} & \colhead{\rev{Required measurement}} & \colhead{\rev{Instruments}} & \colhead{\rev{Detection thresholds}} & \colhead{\rev{Retrieval methods}}
}
\startdata
 & General \rev{3D} shape & NFOV, WFOV & \rev{$\Delta l<$ \SI{30}{\m}} & \rev{Stereophotogrammetry \& photoclinometry} \\
 & \rev{Orbit state} & \rev{-}  & \rev{-} & \rev{-}\\
& \rev{Spin state} & \rev{NFOV}  & \rev{$\Delta l < \omega R_{\rm ast} \Delta t / 2$} & \rev{Landmark identification \& tracking}\\
\shortstack[l]{Origin and evolution\\of Apophis} & Surface craters, boulders, regolith & NFOV  & \rev{$\Delta l<$ \SI{0.2}{\m}} & \rev{Shape analysis}\\
 & Surface \& \rev{exposed} subsurface composition & NFOV & \rev{$\Delta l \lesssim $ \SI{0.1}{\m}} & \rev{Exposed-material imaging} \\
& & \rev{MSIs} & \rev{$\Delta l \lesssim$ \SI{5}{\m}, $\Delta \lambda < $ \SI{50}{\nm}} & \rev{Band-center/depth analysis} \\
 & Interior mass distribution & \rev{-}  & \rev{-} & \rev{Not directly retrieved}\\
 \hline
 & Surface mass flow & NFOV  & \rev{$\Delta l <$ \SI{1}{\m}, $\Delta t >$ \SI{20}{\s}} & \rev{Boulder/regolith tracking}\\
 & & \rev{MSIs}  & \rev{$\Delta l \lesssim$ \SI{5}{\m}, $\Delta \lambda < $ \SI{50}{\nm}} & \rev{Exposed-material spectroscopy}\\
 & Surface spectral change & \rev{MSIs}  & \rev{$\Delta l \lesssim$ \SI{5}{\m}, $\Delta \lambda < $ \SI{50}{\nm}} & \rev{Band-center/depth comparison}\\
\shortstack[l]{Tidal response of Apophis\\during Earth encounter} & Seismic activity & -  & \rev{-} & \rev{-}\\
 & Axial length change & -  & \rev{-} & \rev{-}\\
 & \rev{Spin change rate} & \rev{NFOV} & \rev{$\Delta l < \omega R_{\rm ast} \Delta t / 2$} & \rev{Landmark identification \& tracking}\\
 & \rev{Orbit change rate} & \rev{-} & \rev{-} & \rev{-}\\
\enddata
\tablecomments{\rev{$\Delta l$ refers to spatial resolution, $\Delta t$ refers to required timespan, and $\Delta \lambda$ refers to spectral resolution. A dash indicates that the corresponding quantity is not directly measurable by START. Spin-state and spin-change-rate retrievals with the NFOV camera remain challenging, and inferences about the internal structure would be subject to greater uncertainty.}}
\end{deluxetable*}

\rev{Table~\ref{tab:sci_obj} lists observation-level detection criteria; the resulting measurement uncertainties also depend on the actual observing geometry, instrument calibration, and retrieval methods.}

\begin{deluxetable}{ll}[htb!]
\tablecaption{Gpixel GMAX3265 CMOS Technical Specifications}
\label{tab:gmax3265}
\tabletypesize{\scriptsize}
\tablehead{
\colhead{Parameter} & \colhead{Value}
}
\startdata
Pixel size (\si{\micro\m})  & 3.2$\times$3.2  \\
Pixel count  & 9344$\times$7000 \\
Pixel count (NFOV \rev{and WFOV} usage) & \rev{4800$\times$4800} \\
Bit depth  & 12~bit  \\
% Readout noise (e$^{-}$)  & 1.9 (PGA gain $\times$6)\\&5.0 (PGA gain $\times$1.25) \\
Readout noise (e$^{-}$)  & 5.0 (PGA gain $\times$1.25) \\
Dark current (e$^{-}$/s)  & 5.3 @ die temperature \SI{40}{\celsius}  \\
Peak quantum efficiency $\times$ Fill factor & 65\%  \\
Dynamic range (dB)  & 66 (PGA gain$\times$1.25)  \\
Power consumption (W)  & $<$2.5  \\
Full well capacity (ke$^{-}$)  & 10.0 (PGA gain$\times$1.25)  \\
\enddata
\tablecomments{PGA denotes programmable-gain amplifier.}
\end{deluxetable}

\begin{deluxetable*}{lrrrrr}[htb!]
\tablecaption{Scientific Instruments Parameters\label{tab:inst_params}}
\tablewidth{0pt} % natural width
\tabletypesize{\scriptsize}
\tablehead{
\colhead{} & \colhead{NFOV Camera} & \colhead{WFOV Camera} & \colhead{\rev{MSI-I}} & \colhead{\rev{MSI-II}} & \colhead{\rev{MSI-III}}
}
\startdata
Focal length (\si{\mm}) & 1600 & 160 & \rev{160} & \rev{160} & \rev{160}\\
Aperture (\si{\mm}) & \rev{270} & 32 & 32 & \rev{32} & \rev{32} \\
Pixel count & \rev{4800$\times$4800} & \rev{4800$\times$4800} & \rev{640$\times$512} & \rev{640$\times$512}  & \rev{640$\times$512}\\
FOV (\si{\degree}) & \rev{0.55} & \rev{5.5} & \rev{3.44$\times$2.75} & \rev{3.44$\times$2.75} & \rev{3.44$\times$2.75}\\
iFOV (\si{\micro\radian}/px) & 2 & 20 & \rev{93.75} & \rev{93.75} & \rev{93.75} \\
Spectral range (\si{\nm}) & 450-700 & 450-700 & \rev{700-920} & \rev{900-1200} & \rev{1150-1650}\\
Spectral resolution (\si{\nm}) & - & - & 10 & \rev{20} & \rev{40}\\
Exposure range (\si{\milli\s}) & \rev{0.373\mbox{--}1.18} & 0.266\mbox{--}0.840 & \rev{0.611\mbox{--}6.37} & \rev{0.403\mbox{--}3.53} & \rev{0.435\mbox{--}2.96}\\
Exposure time (\si{\milli\s}) & \rev{0.95} & 0.7 & \rev{5.10} & \rev{2.82}  & \rev{2.37}\\
SNR (\si{\dB})$^a$ & 39.1 & 39.2 & \rev{44.2\mbox{--}45.1} & \rev{43.4\mbox{--}45.0} & \rev{42.4\mbox{--}45.1}\\
\rev{Max tracking rate error} (\si{\degree\per\s}) & \rev{0.036} & \rev{0.49} & \rev{0.32} & \rev{0.57} & \rev{0.68}\\
\enddata
\tablecomments{\rev{$^a$The SNRs of MSIs are values across different spectral channels. Assumed parameters: full well capacity = 40 ke$^{-}$ for MSIs; transmission efficiency = 0.9 for all instruments.}}
\end{deluxetable*}

The detailed requirements include:
(1) \rev{The observation interval should be timed to capture Apophis's expected tidal responses during its approach to ECA and should be as long as practicable.} \rev{The minimum observation duration should be sufficient to resolve a displacement of at least two pixels at a representative speed.} For example, assuming a mass flow speed of $v=$ \SI[per-mode = symbol]{0.1}{\m\per\s} (half of Apophis's surface escape velocity) \rev{and} a pixel resolution of $\Delta l=$~\SI{1}{\m}/px, the required timespan should be \rev{$\Delta t>2 \Delta l/v=$ \SI{20}{\s}}.
Since the tidal response can occur or stop at any time during ECA, a reference temporal resolution of \SI{1}{\s} will help to trace the evolution process and record the event time.
(2) In order to characterize Apophis's detailed surface features such as boulders and craters of no less than \SI{1}{\m}, a spatial resolution of \SI{0.2}{\m}/px is required for the optical camera. \rev{A spatial resolution of approximately \SI{0.05}{\m}/px and a temporal sampling interval of approximately \SI{1}{\s} are considered desirable for detecting the tidally induced regolith motion predicted by \citet{kim2023tidal}.} Determining the spin state of Apophis is more difficult, since it requires a resolution of \rev{$\Delta l < \omega R_{\rm ast} \Delta t / 2$}, \rev{where $\omega \approx \SI[per-mode=symbol]{5.88e-5}{\radian\per\s}$ is the angular velocity of Apophis's spin \citep{pravec2014tumbling} and $R_{\text{ast}} \approx $~\SI{170}{\m} is the radius of the asteroid \citep{brozovic2018goldstone}}, e.g., a spatial resolution of \rev{$\Delta l=$~\SI{0.1}{\m}/px} with a timespan of \rev{$\Delta t=$~\SI{20}{\s}}.
\rev{Such spin-state measurements may also help estimate Apophis's center-of-mass location, thereby constraining its internal mass distribution.}
(3) The spectral measurement should \rev{at least} cover the typical \rev{near-infrared wavelength} range from \rev{\SI{750}{\nm} to \SI{1500}{\nm}}. \rev{A spectral resolution of \SI{50}{\nm} or finer is desired to resolve the center and depth of the principal silicate absorption feature (band I, near \SI{1000}{\nm}) \citep{reddy2018surface}, which provides information on material composition and space-weathering state \citep{binzel2009spectral}.}
(4) For good scientific image quality, the observing geometry should have a phase angle $<$ \SI{60}{\degree}    \citep{atchison2021trajectory}, with a signal-to-noise ratio (SNR) of at least \SI{35}{\dB} for the instruments    \citep{challa2020trajectory, reuter2018flight}. 

The science traceability matrix is presented in Table~\ref{tab:sci_obj}, with a high-resolution NFOV camera, a WFOV navigation camera, and \rev{3 multispectral imagers}. Five kilograms of the spacecraft mass budget is reserved to accommodate potential additional scientific payloads. 

NFOV and WFOV cameras both utilize Gpixel GMAX3265—a back-illuminated, small-pixel scientific CMOS image sensor chosen to achieve high angular resolution and low noise. The technical specifications of this sensor are summarized in Table~\ref{tab:gmax3265}. \rev{Within the CA \SI{\pm10}{\s} window, the relative angular velocity and angular acceleration increase drastically, reaching maximum values of \SI{1.25}{rad/s} and \SI{1.01}{rad/s^2} respectively. Consequently, owing to the finite image-processing and control update rates, the tracking mechanism cannot maintain stable tracking through CA and reaches its tracking limit at some point within the final \SI{10}{\s} before CA. Here, the tracking limit is defined as the point at which the tracking-rate error reaches the modulation transfer function (MTF) threshold of the NFOV camera (see Section~\ref{3.4} for details).} \rev{To guarantee a minimum 20-second observation timespan with a resolution better than \SI{1}{\m}/px before the tracking mechanism reaches this limit, we establish the payload requirements using a conservative worst-case assumption. Assuming that this limit is reached at \SI{-10}{\s} relative to CA, the resolution of \SI{1}{\m}/px must therefore be met before \SI{-30}{\s} relative to CA, which requires the NFOV camera to have an iFOV of less than \SI{3.8}{\micro\radian}/px.} Subject to mass and cost constraints, the NFOV camera’s iFOV is set to \SI{2}{\micro\radian}/px, achieved using a Ritchey–Chrétien imaging system with a \SI{1.6}{\m} focal length. Furthermore, the \rev{MSIs} cover the required spectral range of \rev{750--\SI{1500}{\nm}}, and \rev{InGaAs} image sensors with larger pixel size of \rev{\SI{15}{\micro\m}\,$\times$\,\SI{15}{\micro\m}} were selected to reach higher SNR. The detailed instrument parameters are summarized in Table~\ref{tab:inst_params}.

\subsection{Observation strategy}\label{3.4}
\begin{figure*}[htb!]
    \centering
    \includegraphics[width=0.75\linewidth]{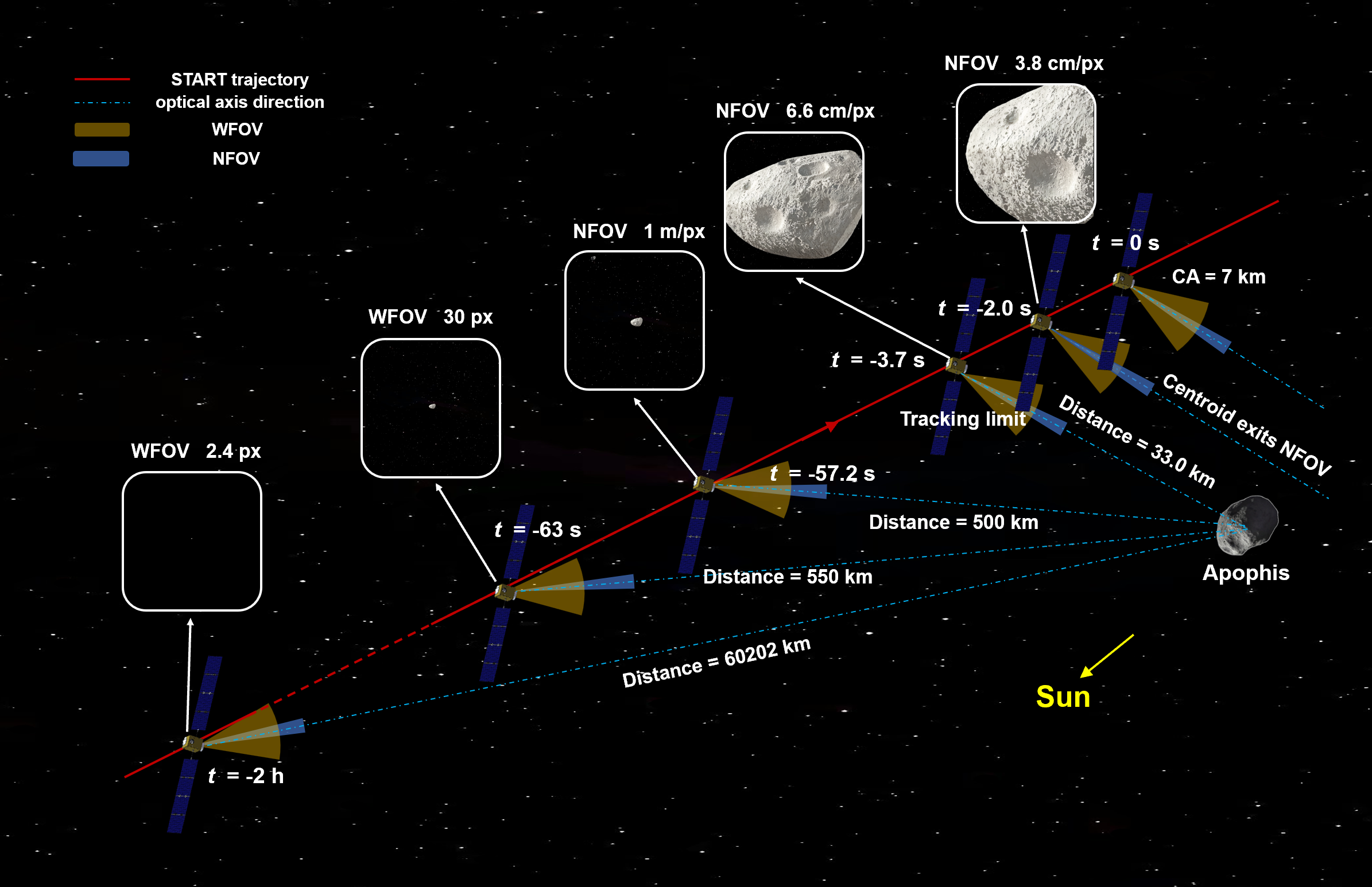}
    \caption{Illustration of the flyby observation process at the CA of \SI{7}{\km}}
    \label{fig:observation}
\end{figure*}

Based on the mission orbit (inclination = \SI{55}{\degree}, right ascension of the ascending node (RAAN) = \SI{102.5}{\degree}, and a CA of \SI{7}{\km}; see Section~\ref{4.1} for details), the relative motion and illumination conditions are analyzed. The calculated relative velocity at CA is \SI[per-mode = symbol]{8.74}{\km\per\s} and remains nearly constant during the observation window. An illustration of the flyby observation process is shown in Figure~\ref{fig:observation}. 

\begin{figure}[htb!]
    \centering
    \includegraphics[width=0.99\linewidth]{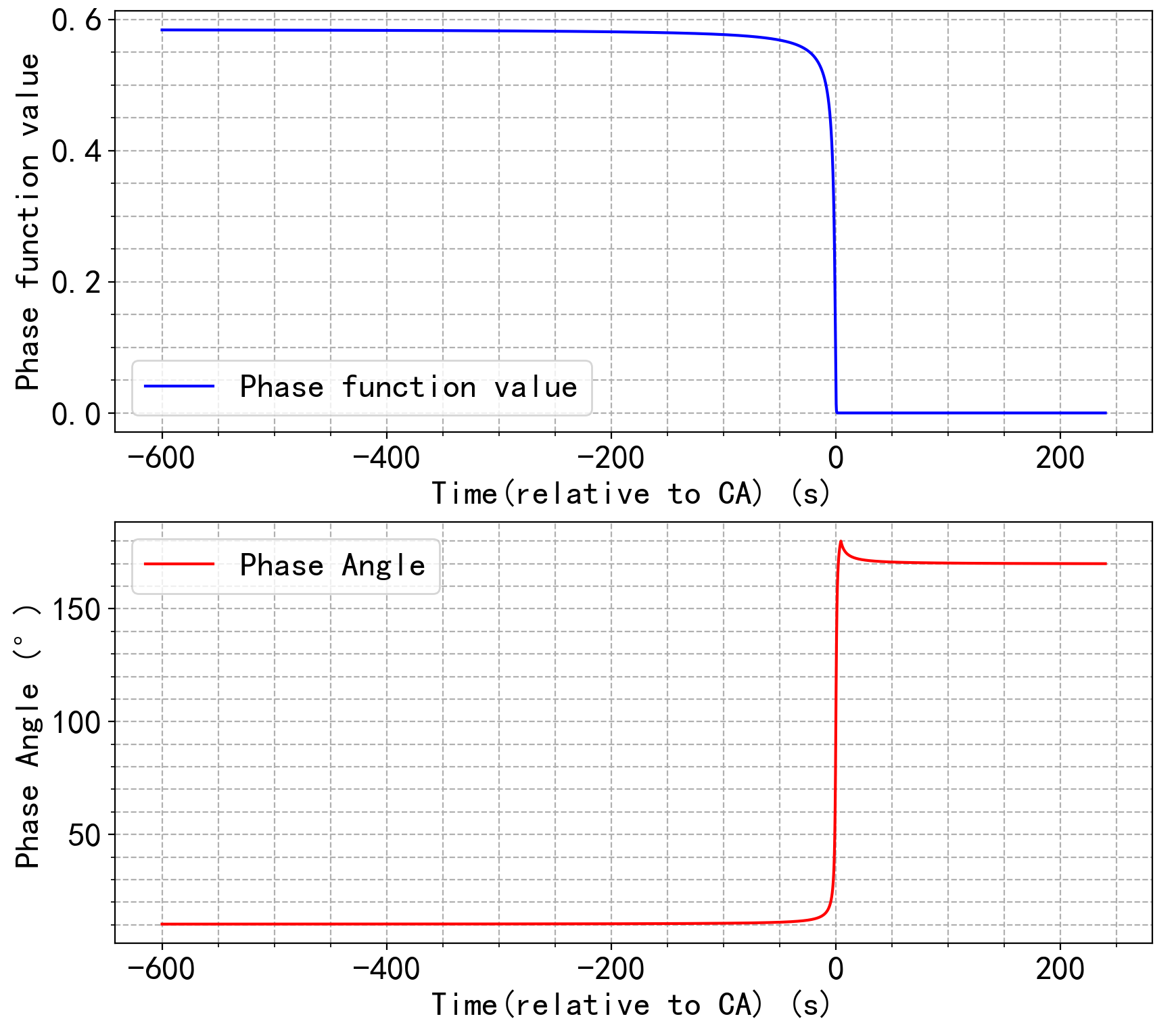}
    \caption{\rev{Phase angle and phase function value evolution.}}
    \label{fig:phase}
\end{figure}

\rev{The phase-angle evolution is shown in Figure~\ref{fig:phase}.} The phase angle remains relatively stable at approximately \rev{\SI{11.0}{\degree}} until about \SI{-60}{\s} relative to CA. From \SI{-60}{\s} to a few seconds prior to CA, the phase angle increases gradually, followed by a sharp surge to approximately \SI{170}{\degree} during the encounter. Consequently, it is a practical approach to utilize the phase angle at \SI{-60}{\s} to calculate the phase function. This allows us to estimate the incoming luminous flux and constrain the camera exposure time to effectively prevent overexposure. Using the H-G photometric model \citep{bowell1989photometric}, the phase function value at \SI{-60}{\s} is calculated to be \rev{0.565}.  

The exposure time for the instruments is bounded by two critical constraints: the lower bound is dictated by the minimum SNR requirements of \SI{35}{\dB}, and the upper bound is limited by the full well capacity (10\,ke$^{-}$ for NFOV/WFOV; \rev{assumed 40\,ke$^{-}$ for MSIs}). To ensure a sufficient dynamic range and prevent pixel saturation under varying illumination, the exposure times are selected so that the expected signal reaches approximately 80\% of the full-well capacity. Table~\ref{tab:inst_params} summarizes the feasible exposure time ranges, the selected exposure time, and the corresponding achievable SNR. 

\rev{MTF characterizes optical imaging quality by measuring contrast retention at various spatial frequencies. Relative motion between the spacecraft and the target induces motion blur, degrading the dynamic MTF from its static baseline. To quantify this degradation, the relative motion is decomposed into tangential and radial components. 
Let $f$ denote the spatial frequency (typically evaluated at the Nyquist frequency, $0.5\,\text{cycles/px}$, which equals half the sampling frequency and represents the highest theoretically reconstructable spatial frequency), and $d$ denote the equivalent relative displacement measured in pixels. The relative MTF under tangential motion blur is expressed as \citep{boreman2001modulation}:}
\begin{equation}
    \rev{\text{MTF}_{\text{tan}} = |\text{sinc}(\pi f d)| = \left| \frac{\sin(\pi f d)}{\pi f d} \right|}
\end{equation}

\rev{Conversely, the relative MTF under radial motion blur is given by \citep{goodman2017introduction}:}
\begin{equation}
    \rev{\text{MTF}_{\text{rad}} = \left| \frac{2J_1(\pi f D)}{\pi f D} \right|, \quad D = \frac{d}{N}}
\end{equation}
\rev{where $N$ is the F-number of the optical system, and $J_1$ is the first-order Bessel function of the first kind.}
\rev{
Given that the payload F-numbers ($N$) range from 5 to approximately 6, the MTF loss induced by radial blur \citep{wertz2011space} is significantly smaller than that of tangential blur \citep{wertz2011space} for an identical $d$. During the staring observation phase, calculations demonstrate that the impact of radial blur remains negligible compared to the empirical engineering threshold of a $0.3\text{\,px}$ tangential blur \citep{wertz2011space}. 
Consequently, tangential blur is identified as the dominant cause of dynamic MTF loss. Adopting the empirical constraint of tangential blur $< 0.3\text{\,px}$ (which corresponds to an MTF loss $< 3.7\%$) as the threshold for stable imaging, the maximum allowable tracking angular velocity errors for each payload are derived. The upper limits of these errors are summarized in Table~\ref{tab:inst_params}.}

During the approach phase, prior to \SI{-10}{\s} (relative to CA), the relative angular velocity and angular acceleration are notably small, providing favorable conditions for the satellite to perform stable staring observations of Apophis. However, after \SI{-10}{\s}, both the angular velocity and acceleration increase rapidly. %转台概述原因，机动能力采用值（30%）, 算法（切换）, 稳定度、指向精度.
\rev{
Consequently, a tracking mechanism is employed to provide greater agility than spacecraft body maneuvering, enabling closer, higher-resolution imaging. The tracking mechanism possesses exceptional capability, with maximum angular acceleration and velocity exceeding \SI[per-mode=symbol]{20}{\degree\per\s\squared} and \SI[per-mode=symbol]{20}{\degree\per\s}, respectively. With low azimuth and pitch inertias ($<0.8$ and $<0.3$~\si{\kg\m\squared}), it minimizes disturbances to the spacecraft. Conservative operational limits of \SI[per-mode=symbol]{5}{\degree\per\s\squared} and \SI[per-mode=symbol]{20}{\degree\per\s} are adopted, which consume $<30\%$ of the maximum torque and $<3\%$ of the momentum capacities of the reaction wheels, leaving ample margins for attitude control.
}

\rev{
For autonomous tracking, real-time WFOV images are processed to compute relative line-of-sight kinematics, dynamically switching between centroid and limb-scanning algorithms based on apparent diameter. Simulations show that, limited by processing and control frequencies, the tracking rate error reaches the NFOV's MTF threshold at \SI{-3.7}{\s}---before reaching the $5^\circ/\text{s}^2$ acceleration limit. Consequently, Apophis's centroid gradually drifts out of the fields of view of the NFOV, WFOV, and MSIs. The simulated single-axis stability ($3\sigma$) is $<0.0015^\circ/\text{s}$ before \SI{-12}{\s}, and $<0.006^\circ/\text{s}$ before \SI{-6.5}{\s}. Correspondingly, the pointing accuracy ($3\sigma$) is $<0.007^\circ$ and $<0.013^\circ$ during these same intervals. These values satisfy the mission requirements over the corresponding intervals.}
\rev{The total pointing error combines the spacecraft's relatively stable attitude control errors (see Section~\ref{5.1} for details) and the tracking mechanism errors; while both are comparable before \SI{-12}{\s}, the latter grows rapidly to dominate during the highest-resolution closest approach.}

Based on the preceding analyses of relative motion, illumination conditions, exposure time calculations, \rev{MTF thresholds and the performance of the tracking mechanism}, the observation schedule has been established. Table~\ref{tab:timeline} summarizes the observation schedule, detailing key observation events—such as instrument power-on, reaching specified thresholds for resolution or \rev{tangential} blur—along with their corresponding time and distance. 

\begin{figure*}[htb!]
    \centering
    \includegraphics[width=0.8\linewidth]{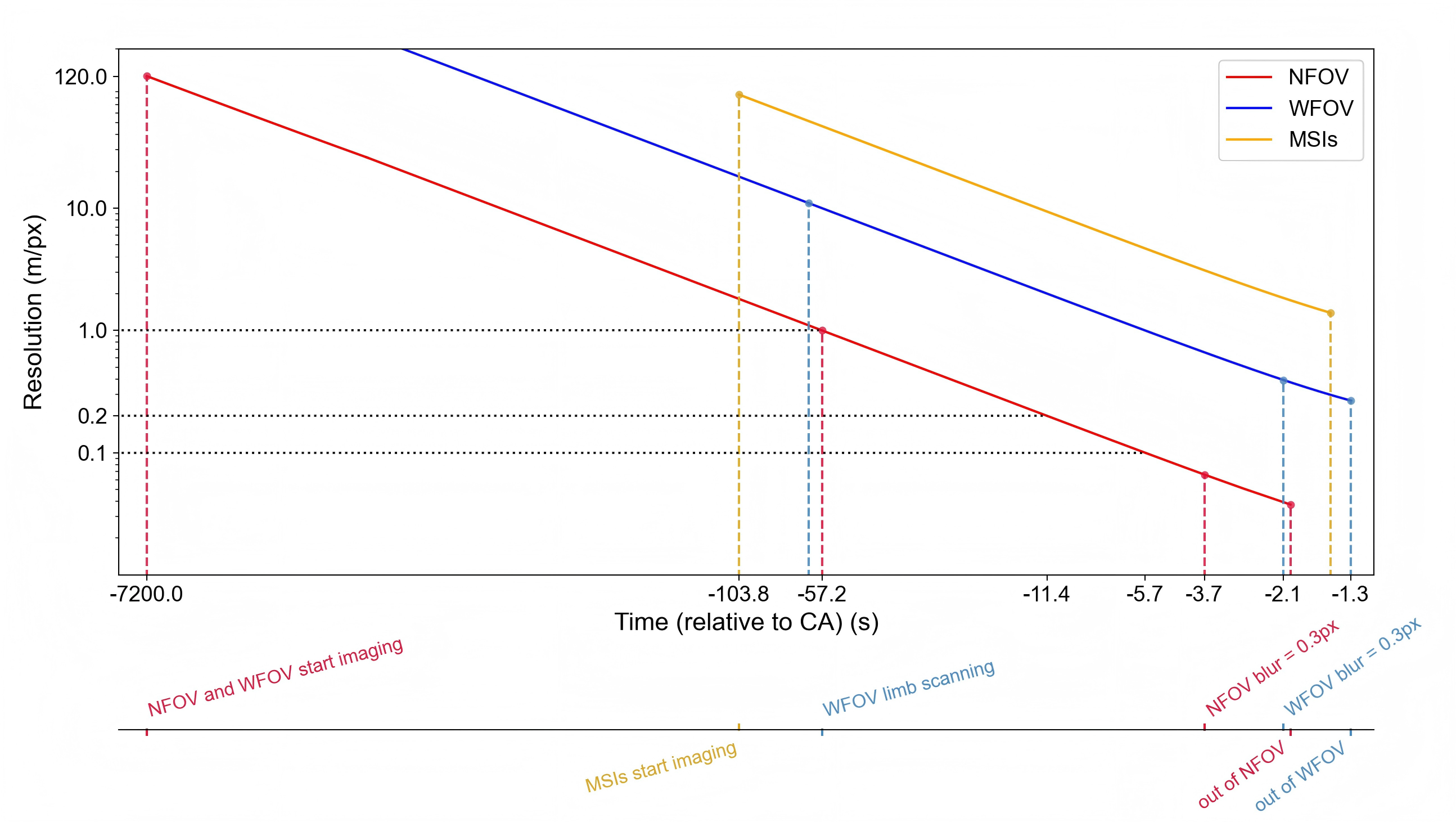}
    \caption{\rev{Resolution of the NFOV, WFOV and MSIs. Solid lines depict changes in resolution of the NFOV camera, WFOV camera and MSIs during the flyby observation process. Dashed lines mark the timing of key observation events—including instrument power‑on, reaching specified thresholds for resolution or tangential blur. Dotted lines indicate key resolution values during the observation. Note that the data are presented on a log-log scale.}}
    \label{fig:resolution}
\end{figure*}

Regarding the overall observation performance, the NFOV camera can achieve an observation timespan of $>$~\SI{50}{\s} with a spatial resolution better than \SI{1}{\m}/px and a \rev{tangential} blur of $<$~0.3\,px. \rev{Under the 0.3-pixel blur constraint, the NFOV camera reaches a best usable resolution of \SI{6.6}{\cm/px}, approaching the desired resolution of \SI{5}{\cm/px}.} Similarly, the WFOV camera achieves an optimal resolution of \rev{\SI{0.39}{\m/px}} under \rev{a tangential blur constraint of 0.3\,px}. Meanwhile, the \rev{MSIs} provide an optimal resolution of \rev{\SI{1.9}{\m/px}} with a \rev{tangential} blur of $<$~0.3\,px. The continuous variation in resolution of NFOV, WFOV, and \rev{MSIs} throughout the flyby is illustrated in Figure~\ref{fig:resolution}. 

% 为了实现科学目标而制定的载荷性能和观测策略（何时开机，效果如何）
% 图片参照群里发的Lucy图片

\begin{deluxetable*}{llll}[htb!]
\tablecaption{Mission observation schedule \label{tab:timeline}}
\tablewidth{0pt}
\tabletypesize{\scriptsize}
\tablehead{
  \colhead{Distance (\si{\km})} &
  \colhead{\parbox[t]{2cm}{\raggedright Time (\si{\s}) \\ (relative to CA)}} &
  \colhead{Timeline Events} &
  \colhead{Notes}
}
\startdata
60202 & $-$7200 & \parbox[t]{6cm}{\raggedright NFOV and WFOV start imaging. \\ \rev{WFOV target acquisition and tracking (centroid algorithm).}\vspace{3pt}} & \parbox[t]{5.5cm}{\raggedright Apophis apparent magnitude: 2.66; \\Apophis apparent diameter (\rev{WFOV}): \rev{2.4}\,px;\\Suitable for centroid algorithm.\vspace{3pt}} \\ 
\rev{906.7} & \rev{$-$103.8} & \parbox[t]{6cm}{\raggedright \rev{MSIs start imaging.}\vspace{3pt}} & \parbox[t]{5.5cm}{\raggedright \rev{Apophis covers 4$\times$4 spectral channels.}\vspace{3pt}} \\ 
\rev{550} & \rev{$-$63.0} & \parbox[t]{6cm}{\raggedright \rev{WFOV target tracking (limb scanning algorithm).}\vspace{3pt}} & \parbox[t]{5.5cm}{\raggedright \rev{Apophis apparent diameter (WFOV): 30\,px;\\Meets limb scanning requirements.}\vspace{3pt}} \\ 
500 & $-$57.2 & \parbox[t]{6cm}{\raggedright NFOV resolution: 1\,m/px\vspace{3pt}} & \parbox[t]{5.5cm}{\raggedright WFOV resolution: 10\,m/px;\\\rev{High-resolution imaging window begins;}\\ \rev{MSIs resolution: 46.9\,m/px.}\vspace{3pt}} \\ 
\rev{33.0} & \rev{$-$3.7} & \parbox[t]{6cm}{\raggedright \rev{NFOV tangential blur: 0.3\,px.\\Tracking mechanism hits the tracking limit.}\vspace{3pt}} & \parbox[t]{5.5cm}{\raggedright \rev{NFOV resolution: 6.6\,cm/px.}\vspace{3pt}} \\ 
\rev{20.5} & \rev{$-$2.2} & \parbox[t]{6cm}{\raggedright \rev{MSI-I tangential blur: 0.3\,px.}\vspace{3pt}} & \parbox[t]{5.5cm}{\raggedright \rev{MSIs resolution: 1.92\,m/px.}\vspace{3pt}} \\ 
\rev{19.6} & \rev{$-$2.1} & \parbox[t]{6cm}{\raggedright \rev{WFOV tangential blur: 0.3\,px.}\vspace{3pt}} & \parbox[t]{5.5cm}{\raggedright \rev{WFOV resolution: 0.39\,m/px;\\MSI-II tangential blur: 0.35\,px;\\MSI-III tangential blur: 0.42\,px.}\vspace{3pt}} \\ 
\rev{18.8} & \rev{$-$2.0} & \parbox[t]{6cm}{\raggedright \rev{Image centroid of Apophis exits the NFOV.}\vspace{3pt}} & \parbox[t]{5.5cm}{\raggedright \rev{NFOV resolution: 3.8\,cm/px;\\Tangential blur: 6.8\,px.}\vspace{3pt}} \\ 
\rev{14.9} & \rev{$-$1.5} & \parbox[t]{6cm}{\raggedright \rev{Image centroid of Apophis exits the FOV of MSIs.}\vspace{3pt}} & \parbox[t]{5.5cm}{\raggedright \rev{MSIs resolution: 1.40\,m/px;\\MSI-I tangential blur: 4.1\,px.}\vspace{3pt}} \\ 
\rev{13.3} & \rev{$-$1.3} & \parbox[t]{6cm}{\raggedright \rev{Image centroid of Apophis exits the WFOV.}\vspace{3pt}} & \parbox[t]{5.5cm}{\raggedright \rev{WFOV resolution: 0.27\,m/px;\\Tangential blur: 4.4\,px.}\vspace{3pt}} \\ 
\rev{7} & \rev{0} & \parbox[t]{6cm}{\raggedright \rev{Payloads off; tracking mechanism decelerates.}\vspace{3pt}} & \parbox[t]{5.5cm}{\raggedright \rev{CA (approximately coincident with ECA)}\vspace{3pt}} \\
\enddata
\tablecomments{\rev{If Apophis's geometric apparent diameter is smaller than the Airy disk, the apparent diameter is taken as the sum of the geometric diameter and the Airy disk diameter.}}
\end{deluxetable*}

\section{Trajectory design}

\begin{figure*}[ht!]
    \centering
    \includegraphics[width=0.8\linewidth]{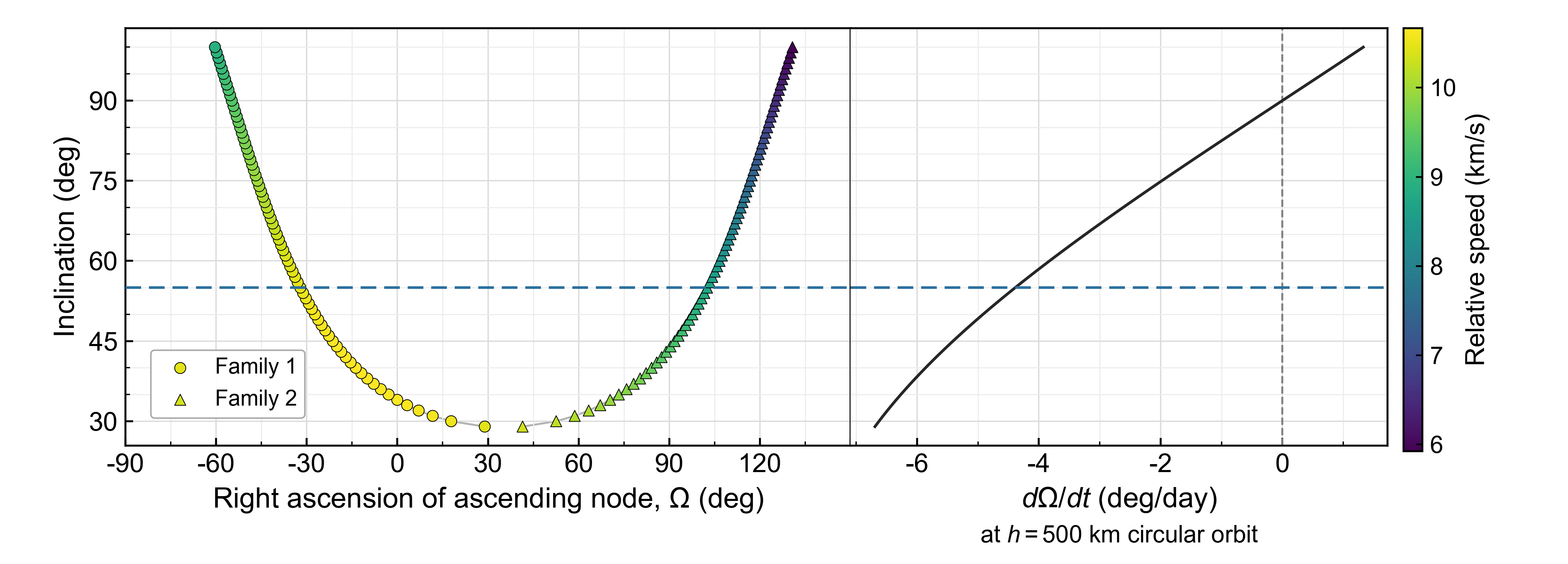}
    \caption{Feasible interception solutions: the left panel shows geometry ($\Omega$ vs. inclination) colored by relative velocity, while the right panel plots the nodal precession rate ($\dot{\Omega}$). The blue dashed line marks the selected \SI{55}{\degree} inclination.}
    \label{fig:5}
\end{figure*}

Targeting Apophis \rev{at its ECA, when its geocentric distance is approximately \SI{38000}{\km}}, START adopts a geocentric interception strategy rather than a typical deep-space rendezvous.
\rev{Centering the flyby near ECA minimizes the required mission-orbit radius; an earlier or later intercept would require a higher orbit, increasing transfer time and propellant consumption.}
This approach eliminates the dependency on heavy launchers or deep-space communications, enabling a viable low-cost commercial mission profile. 
Furthermore, the trajectory serves as a critical demonstration of SEP capabilities for extensive orbit raising from LEO to high Earth orbits.

\begin{figure*}[ht!]
    \centering
    \includegraphics[width=0.9\linewidth]{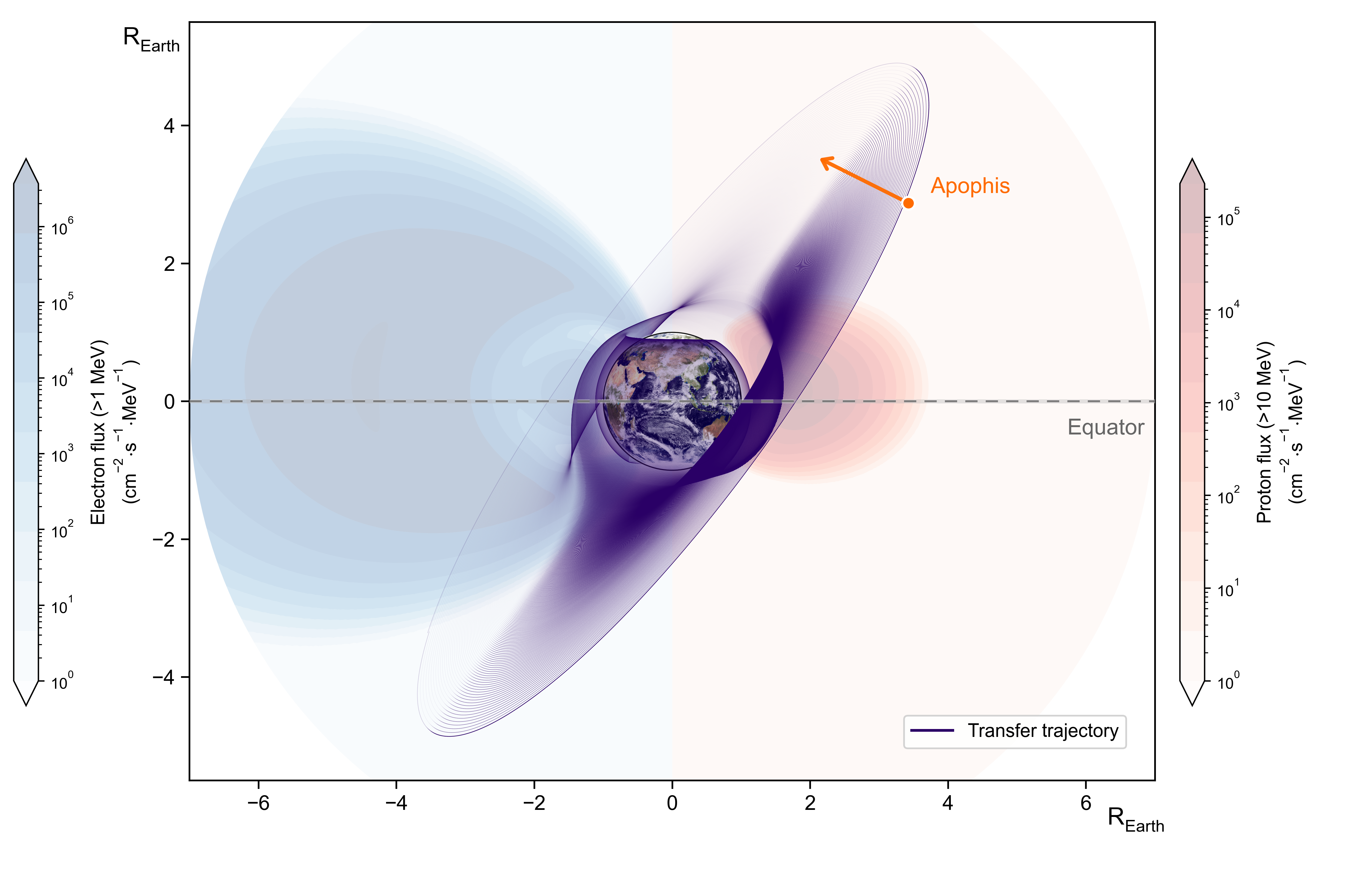}
    \caption{Equatorial view of the 3D spiral trajectory. \rev{The background contours show the modeled electron (blue) and proton (red) flux distributions in Earth's radiation belts.} The orange arrow denotes Apophis at Earth closest approach and its geocentric velocity vector.}
    \label{fig:6}
\end{figure*}

\subsection{Mission orbit} \label{4.1}

At its ECA, Apophis's geocentric declination is approximately \SI{29}{\degree}N, requiring the spacecraft’s orbital inclination to lie between \SI{29}{\degree} and \SI{151}{\degree} \rev{for its orbital plane to pass through the encounter point}. As shown in Figure~\ref{fig:5} (left panel), each feasible inclination yields two families of RAAN solutions, of which the family with the lower relative velocity is selected.

\rev{The high velocity-increment cost of orbital-plane changes makes rideshare injection conditions an additional constraint on mission-orbit design.} Ideally, the spacecraft's inclination should approach Apophis's local geocentric trajectory inclination ($\sim$\SI{140.6}{\degree}) to minimize the relative velocity during the flyby. \rev{However, rideshare opportunities to retrograde orbits with inclinations above \SI{100}{\degree} are virtually nonexistent.}

\rev{A GTO rideshare was also considered because orbital-plane and apsidal adjustments require relatively low $\Delta v$ at high altitude. A typical GTO, however, has a low inclination, a near-equatorial apogee, and slow nodal precession. Matching the ECA geometry would therefore still require substantial changes in inclination, RAAN, and apsidal orientation, with corresponding penalties in propellant and transfer time. Rideshare opportunities to such highly elliptical orbits are also relatively costly and infrequent.}

\rev{Using a highly elliptical orbit as the mission orbit could reduce the relative velocity by placing the encounter near apogee, but its repeated radiation-belt crossings would be unfavorable for long-term operation and potential extended missions. START therefore adopts an SEP transfer from a \SI{500}{\km} LEO to a high-altitude circular mission orbit.}

After injection into LEO, the spacecraft utilizes Earth's $J_2$ perturbation in the \rev{\SI{500}{\km}} parking orbit to phase its RAAN before initiating orbit raising. A trade-off exists between the flyby velocity and the orbital-plane phasing capability. High-inclination orbits (e.g., polar or Sun-synchronous) offer lower relative velocities (\SIrange[per-mode=symbol]{6}{7}{\km\per\s}) but have slow nodal precession rates, \rev{making the RAAN at the start of orbit raising strongly dependent on the injected orbital plane}. At this altitude with a \SI{55}{\degree} inclination, the Earth's $J_2$ perturbation induces a nodal precession rate (${\rm d}\Omega/{\rm d}t$) of approximately \rev{\SI[per-mode=symbol]{-4.39}{\degree\per\day}}. \rev{A full \SI{360}{\degree} RAAN cycle therefore takes only $\sim$82 days, reducing the dependence on the injected orbital plane and broadening rideshare opportunities within an acceptable parking duration.} We therefore select a nominal inclination of \SI{55}{\degree}. This configuration corresponds to a final RAAN of \SI{102.5}{\degree} and a relative velocity of \SI[per-mode=symbol]{8.74}{\km\per\s} at the CA.

\rev{For the flyby geometry considered here, a smaller CA distance causes the line-of-sight angular rate to reach the tracking limit closer to CA and hence later during approach. This allows stable imaging to continue to shorter ranges and achieve finer spatial resolution. RAMSES's main spacecraft, however, is planned to hover approximately \SI{5}{\km} from Apophis during the close-encounter phase \citep{martino2025operational}. START therefore adopts a nominal CA distance of \SI{7}{\km}, retaining the imaging benefit of a close flyby while providing an operational margin beyond RAMSES's planned hovering distance.}

Finally, \rev{at this nominal CA distance, variations in the local flyby direction have little effect on pre-CA illumination.} Owing to the nearly sunward direction of Apophis's high geocentric velocity at ECA, its sunlit hemisphere remains visible throughout pre-CA observations (see Section~\ref{3.4} for details). \rev{One such geometry would also allow an auxiliary body-mounted camera to image a partial solar occultation by Apophis after CA.}

\begin{figure*}[ht!]
    \centering
    \includegraphics[width=0.9\linewidth]{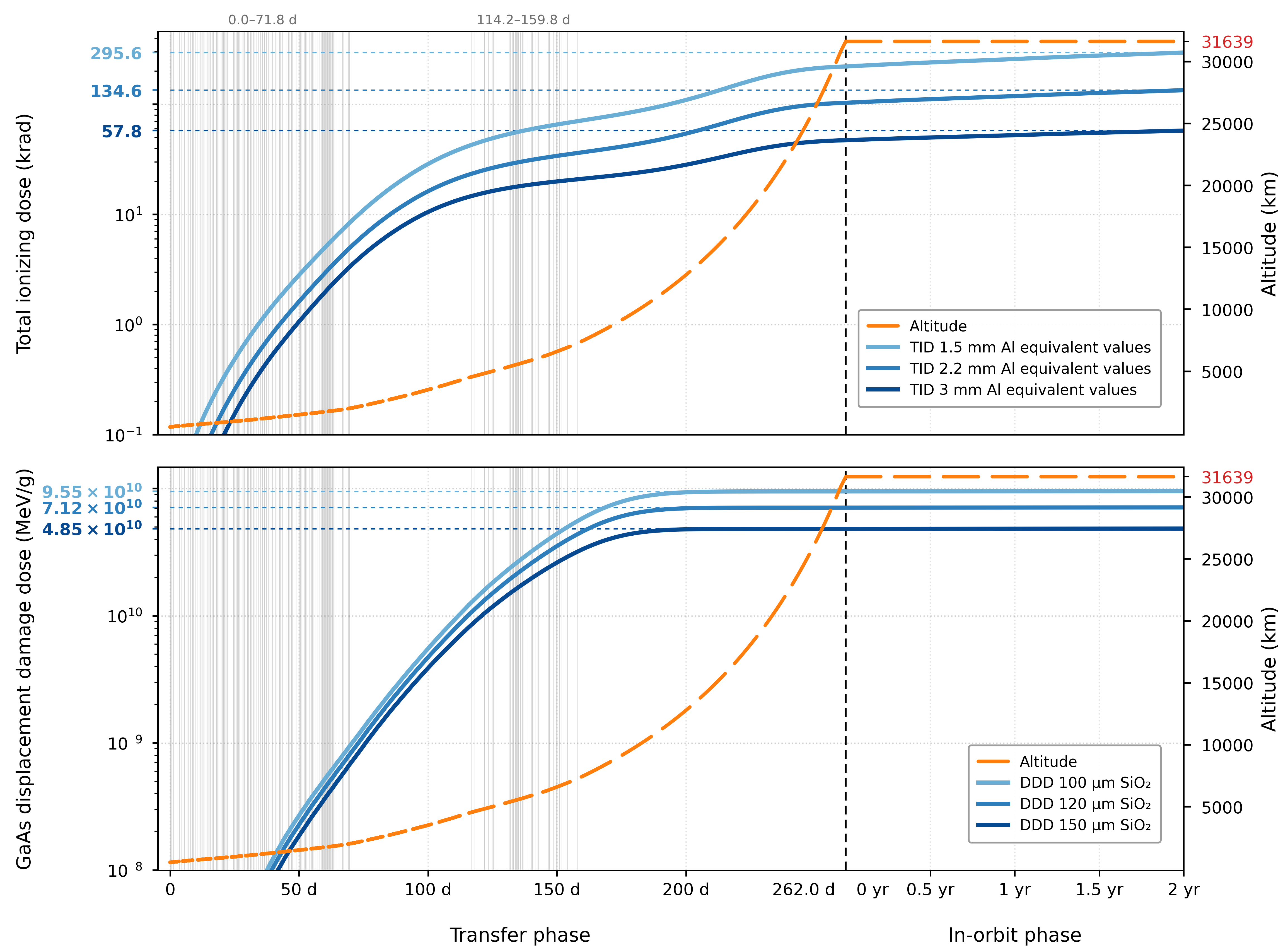}
    \caption{\rev{Evolution of total ionizing dose (TID) and GaAs displacement damage dose (DDD) during the transfer and in-orbit phases. The orange dashed curve shows the spacecraft altitude. Dashed lines indicate the target orbit and transfer completion. The TID is evaluated for aluminum-equivalent shielding thicknesses of \SI{1.5}{\mm}, \SI{2.2}{\mm}, and \SI{3.0}{\mm}, and the DDD is evaluated for SiO$_2$ coverglass thicknesses of \SI{100}{\micro\m}, \SI{120}{\micro\m}, and \SI{150}{\micro\m}. The black dashed line marks transfer completion at approximately 262~days, gray bands indicate eclipse-induced propulsion shutdowns, and horizontal dotted lines denote the cumulative dose levels after two years in the target orbit.}}
    \label{fig:7}
\end{figure*}

\subsection{Low thrust transfer}

SEP has proven effective for orbit raising initiated from elliptical orbits, as demonstrated by 
geostationary telecommunication satellites \citep{Liang2025APSTAR6E,autric2018telecom,poole2007boeing} and the SMART-1 lunar mission \citep{racca2002smart}.
\rev{START, however, raises its orbit directly from a \SI{500}{\km} circular LEO to the selected high-altitude circular mission orbit, entailing a large orbital-energy increase and prolonged radiation exposure within Earth's Van Allen belts. More critically, the compressed timeline of this rapid-response mission precludes extending the transfer merely to reduce propellant consumption.}

The spacecraft employs a xenon-based Hall SEP system delivering a maximum thrust of \SI{60}{\mN} with a power input of $\sim$\SI{1000}{\W}. \rev{Using direct collocation methods \citep{zou2026vectorized}, a minimum-time transfer of approximately 262~days is obtained, with sustained thrust directed predominantly along the instantaneous velocity vector. The trajectory is subsequently refined using a high-precision dynamical model accounting for high-order Earth zonal harmonics, lunisolar perturbations, atmospheric drag, and solar radiation pressure.}

\rev{The operational design maintains active solar tracking for maximum power generation}, while thrust is suspended during Earth eclipses and for a mandatory 15-minute maintenance interval after every 24~hours of continuous operation. \rev{Together with additional shutdowns for eccentricity control, these interruptions lead to nearly 2,000 thruster starts over the transfer. For an initial spacecraft mass of \SI{300}{\kg}, the refined trajectory consumes approximately \SI{71.1}{\kg} of xenon, corresponding to an equivalent velocity increment of \SI[per-mode=symbol]{4.38}{\km\per\s}. The RAAN naturally regresses by approximately \SI{334}{\degree} during the transfer.} The resulting 3D trajectory and \rev{altitude} evolution are presented in Figures~\ref{fig:6} and~\ref{fig:7}, respectively.

\rev{Operational planning supplements this refined trajectory with contingency margin for fault-induced thrust interruptions. An interruption lasting two to three orbital periods would prolong low-altitude residence and produce additional $J_2$-driven RAAN regression. Direct correction through a dedicated plane-change maneuver is impractical owing to its high velocity-increment cost. Accordingly, two adjustable 10-day coast intervals are scheduled, one in the low-flux slot between the inner and outer belts and the other above the outer belt's peak-flux region. The excess regression is instead offset by shortening a later coast interval, thereby reducing the regression accumulated thereafter. Relative to the minimum-time solution, orbit raising is scheduled to begin approximately 20 days earlier to accommodate these intervals. During the final high-altitude leg, the resulting RAAN error is sufficiently small for subsequent trajectory correction.}

\subsection{Space environment}

\rev{As illustrated in Figure~\ref{fig:6}, the prolonged orbit-raising phase exposes the spacecraft to radiation from the proton-dominated inner belt and the electron-dominated outer belt \citep{ginet2013ae9}, thereby making the accumulated total ionizing dose (TID) a major design constraint for COTS electronics. As shown in Figure~\ref{fig:7}, \SI{2.2}{\mm} shielding cannot satisfy the design limit of 80\,krad(Si). Therefore, at least \SI{3.0}{\mm} of aluminum-equivalent shielding is adopted, together with additional local shielding for critical units such as the onboard computer, power electronics, and memory devices \citep{ECSS2025RHA}.}

\rev{Energetic proton and electron irradiation produces displacement defects in GaAs, leading to degradation of carrier transport properties and solar-cell output power. Based on the displacement damage dose analysis \citep{messenger2001modeling}, a \SI{150}{\micro\m} coverglass is adopted, and the resulting radiation-induced degradation is included in the end-of-life power assessment of the solar array \citep{gruginskie2021proton}.}

\rev{Single-event effects of concern mainly include single-event upsets, single-event latch-up (SEL), and single-event functional interrupts. Based on the mission linear energy transfer (LET) integral spectrum, a representative screening threshold of \(37~\mathrm{MeV\,cm^2\,mg^{-1}}\) for non-recoverable single-event effects is adopted for the preliminary assessment \citep{NASA2021AIRES}, indicating relatively low direct heavy-ion latch-up exposure for devices with comparable or higher LET thresholds \citep{Tylka1997CREME,Sexton2003SEE}. Critical avionics are protected using error detection and correction, watchdog timers, state or configuration readback, and fault recovery mechanisms \citep{rudolph2014csp}. Current monitoring and limiting are implemented to prevent SEL-induced overcurrent from causing permanent device damage \citep{Sexton2003SEE}. Autonomous reset or power cycling is also employed to recover equipment from single-event functional interrupts \citep{Ladbury2013SEFI}.}

\rev{Spacecraft surface charging and the associated electrostatic-discharge risks are mitigated through proper grounding and bonding. Equipment enclosures, structural components, cable shields, and conductive surfaces are connected to a common spacecraft structural ground to avoid floating conductors and local charge accumulation \citep{Garrett1981Charging}.}

\subsection{\rev{Flyby navigation and control}}

\rev{Unlike conventional deep-space flybys, START remains in Earth orbit and uses a ground-in-the-loop navigation architecture. High-altitude orbit determination exploits global navigation satellite system (GNSS) signals spilling beyond Earth's limb from satellites on its far side, supplemented by ground-based laser ranging via an onboard retroreflector and scheduled telemetry, tracking, and command (TT\&C) contacts with commercial ground stations in China and overseas. GNSS measurements are downlinked for precise orbit determination and maneuver design; correction commands are then uplinked, and the corrected orbit is verified through subsequent measurements.}

\rev{START will reach the mission orbit at least one month before CA for final checkout, trajectory refinement, and encounter rehearsals. Periodic trajectory corrections, nominally every 10~days, maintain the required flyby geometry. Final orbit determination and correction are completed approximately 5~h before CA, after which the spacecraft follows a ballistic trajectory through the encounter. Onboard optical navigation thereafter supports autonomous target acquisition and line-of-sight tracking for imaging without feeding back into trajectory correction.}

\rev{Propagation of the covariance associated with JPL Solution 220 to the 2029 ECA gives a current $1\sigma$ position uncertainty of approximately \SI{1.2}{\km} for Apophis \citep{Farnocchia2024Apophis220}. This uncertainty is expected to decrease substantially as additional ground-based observations are incorporated before the encounter. For START, orbit-deviation propagation using the adopted post-correction navigation-error model, corresponding to a position uncertainty of \SI{30}{\m} ($1\sigma$ per axis), yields a $3\sigma$ closest-approach-distance dispersion of approximately \SI{0.4}{\km} around the nominal \SI{7}{\km} value. Before final targeting, the flyby geometry will be coordinated with RAMSES and other confirmed missions through the exchange of ephemerides, covariances, and maneuver plans. Together with the above navigation accuracy, the selected geometry provides sufficient separation from Apophis and nearby spacecraft during the flyby.}

\section{Spacecraft design}

As a \rev{\SI{300}{\kg}} small satellite, the START spacecraft is required to \rev{accommodate a \SI{25}{\kg} scientific payload mass allocation}, perform the orbital transfer with a total delta-v exceeding \SI[per-mode=symbol]{4}{\km\per\s}, and complete the flyby investigation of the asteroid Apophis at a relative velocity of up to \SI[per-mode=symbol]{8.74}{\km\per\s}. 
To meet the demands of this challenging mission, the spacecraft will differ from the standard off‑the‑shelf spacecraft bus architectures typically adopted by conventional medium‑Earth‑orbit satellites or optical remote‑sensing satellites. Instead, it must be purpose‑designed to deliver enhanced performance.

\subsection{Attitude determination and control system} \label{5.1}

\rev{To maintain continuous observation during the high-speed encounter, START combines stable spacecraft attitude control with a high-speed tracking mechanism.} \rev{According to the relative motion profile, the tracking mechanism is required to provide an angular acceleration of \SI[per-mode=symbol]{5}{\degree\per\s\squared} and a maximum angular rate of at least \SI[per-mode=symbol]{20}{\degree\per\s}, thereby keeping Apophis within the NFOV at ranges down to \SI{33}{\km} during the high-speed encounter. It must also limit the angular-rate error to \SI[per-mode=symbol]{0.036}{\degree\per\s}, so that tangential motion blur remains below 0.3~px during each \SI{0.95}{\milli\s} exposure.}

The spacecraft employs one \rev{3-axis} fiber optic gyroscope, three star trackers, and three Sun sensors, \rev{providing an attitude-determination accuracy of} \rev{\SI{0.00225}{\degree}} (3$\sigma$) on the mission orbit. The actuator system comprises \rev{4 reaction wheels} and a reaction control system (RCS) equipped with eight thrusters. 
\rev{The four reaction wheels are arranged in a skewed configuration, with each wheel delivering a maximum control torque of at least \SI{0.1}{\N\m} and a maximum angular momentum of no less than \SI{4}{\N\m\s} at 6000 rpm.} \rev{The RCS is capable of generating control torques of 0.85, 0.58, and \SI{0.72}{\N\m} along the three respective axes. During the flyby phase, to accommodate the maneuver of the tracking mechanism, the satellite is required to maintain three-axis stability in the Earth-centered inertial (ECI) frame, achieving a pointing accuracy better than \SI{0.005}{\degree} (3$\sigma$) and an attitude stability better than \SI[per-mode = symbol]{0.002}{\degree\per\s} (3$\sigma$) per axis. Additionally, to detumble the satellite after launch vehicle separation and unload the saturated angular momentum of the reaction wheels, the spacecraft is required to carry at least \SI{15}{\kg} of anhydrous hydrazine.}

\subsection{Propulsion system and power system}

Regarding orbital transfer, the satellite is required to have the capability for continuous orbit-raising for up to one year, a minimum operational lifespan of three years, and sufficient orbit control accuracy. Therefore, it must be equipped with highly reliable propulsion and power systems to support its transfer from a \rev{\SI{500}{\km}-altitude LEO} to \rev{the selected circular mission orbit}. The spacecraft will carry \rev{\SI{80}{\kg}} of xenon propellant. Its two Hall thrusters, operating in relay, will provide a thrust of at least \SI{60}{\mN} while maintaining a specific impulse of approximately \rev{\SI{1650}{\s}}, thereby supporting the spacecraft in completing this long-distance transfer. In addition, a remaining total impulse of approximately \rev{\SI{140}{\kilo\N\s}} is retained to provide the extra propulsion required for orbit control and navigation, to help ensure successful mission execution, and to enable the spacecraft to conduct potential extended missions.

The thrusters will require \SI{1020}{\W} of power, raising the spacecraft’s total power demand to as high as \rev{\SI{1500}{\W}}. To account for power degradation caused by radiation effects and to ensure that the power demands of the thruster can always be met, the solar arrays were designed with a total area of \rev{\SI{10.25}{\m\squared}}. They are driven by a pair of \rev{single-axis} solar array drive assemblies (SADAs). The arrays are equipped with triple-junction gallium arsenide solar cells, which offer a conversion efficiency of 30$\%$. Even at the end of life, these cells can generate a total of \rev{\SI{2297}{\W}} of electrical power while operating at their maximum power point, comfortably satisfying the satellite's total power demand.

\subsection{Configuration and structural design}

\begin{figure*}[htb!]
    \centering
    \includegraphics[width=0.9\linewidth]{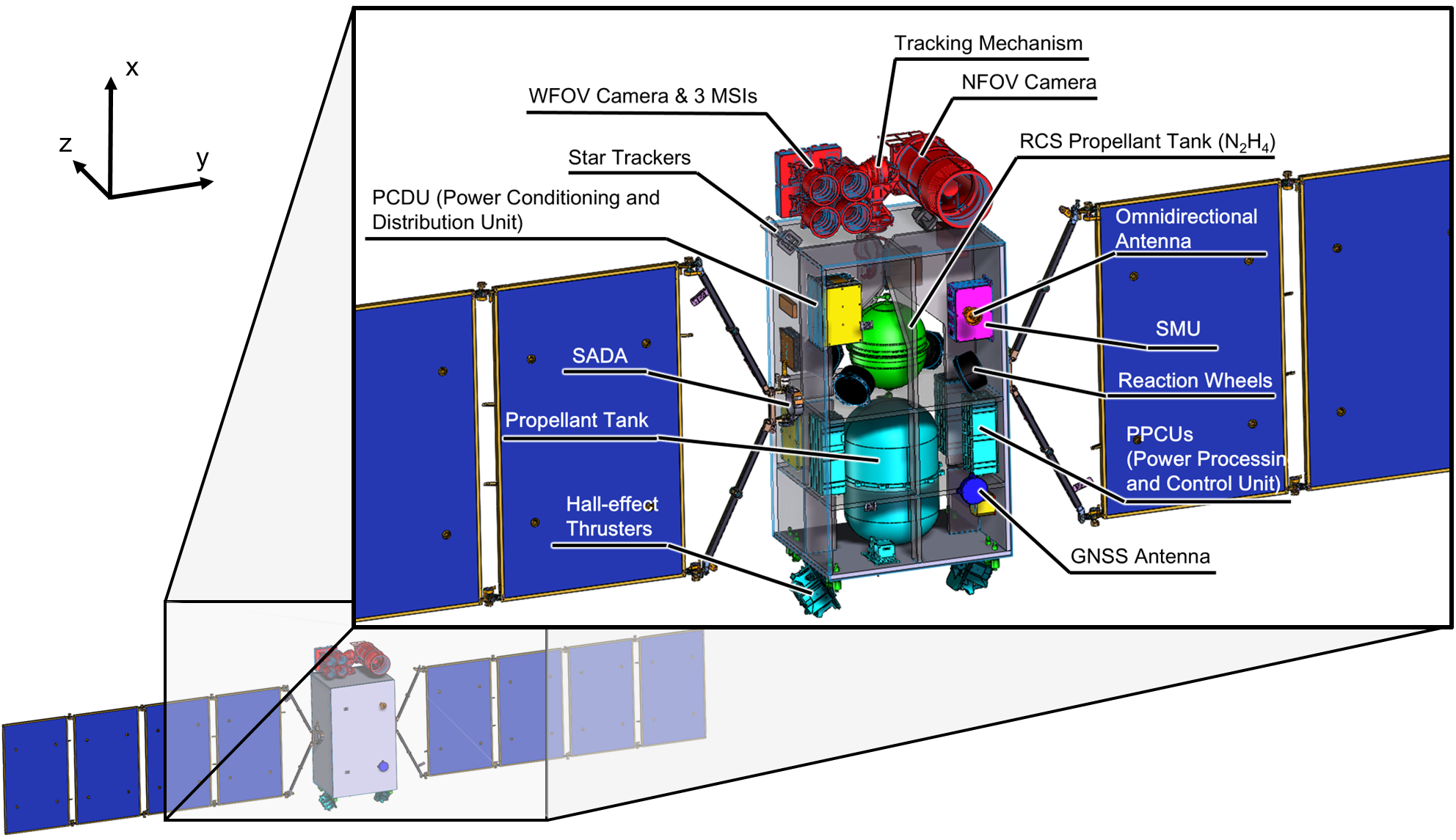}
    \caption{START spacecraft design. Main elements and payloads are indicated.}
    \label{fig:spacecraft}
\end{figure*}

\rev{The spacecraft employs a box-type structure layout that is amenable to low-cost fabrication, with mutually perpendicular side panels and partition panels forming the primary load-bearing structure, yielding a geometrically symmetric configuration, as shown in Figure~\ref{fig:spacecraft}. Two Hall thrusters equipped with two-axis gimbal mechanisms for thrust vector control, together with eight RCS thrusters, are mounted on the $-X$ panel; the remaining propulsion system components are installed on the adjacent side panels. The propellant tanks are arranged along the X‑axis at the center of the satellite, ensuring that the center of mass does not undergo significant displacement as propellant is consumed. By contrast, the scientific payloads—carried by the tracking mechanism—are mounted on the $+X$ panel. The solar arrays deploy from the $\pm Y$ panels, with the rotation axes of their SADAs precisely aligned parallel to the satellite’s Y‑axis. This layout ensures that the angle between the rotation axes of the solar arrays and the thrust direction is always greater than $70\,^\circ$, thereby preventing mutual interference between the satellite’s power generation and propulsion. Other equipment, such as the power processing unit and antennas, are distributed uniformly across the various side and partition panels according to system requirements.}

\subsection{Thermal control system}

\rev{In terms of the thermal environment, the spacecraft is subjected to sustained solar irradiation and deep-space exposure. Given mass constraints, passive thermal control predominates, employing multilayer insulation blankets and thermal control coatings. Heat pipes are strategically integrated throughout the structure to minimize temperature gradients and ensure thermal homogenization. The electronic equipment within the satellite compartment (e.g., the satellite management unit (SMU)) imposes stringent temperature range requirements of \SI{-15}{\celsius} to \SI{45}{\celsius}. Thermal model analysis of the satellite indicates that the aforementioned thermal control measures can successfully maintain the temperature range of the internal electronic equipment between \SI{-14.4}{\celsius} and \SI{26.8}{\celsius}.}

\subsection{\rev{Communication systems}}

\rev{The satellite data transmission system and the TT\&C system both operate in the X band at \SI{8}{\giga\Hz}. A lightweight phased-array antenna is used for science data downlink, whereas an omnidirectional antenna supports telemetry, tracking, and command.}

\rev{The phased-array antenna achieves an effective isotropic radiated power (EIRP) of 30\,dBW. In conjunction with a \SI{7.3}{\m}-aperture ground station, the system supports a data rate of at least 20\,Mbps and maintains a link margin of \SI{5.3}{\dB} to ensure reliable communications. With multiple ground stations located in mainland China, Africa, and South America, the complete mission dataset can be downlinked in as little as one week.}

\rev{The omnidirectional antenna has an EIRP of 0\,dBW and a minimum gain of $-1$\,dBi. In conjunction with a \SI{7.3}{\m}-aperture ground station, the TT\&C system supports uplink and downlink data rates of 2000 and 4096\,bps, respectively, with corresponding link margins of \SI{3.17}{\dB} and \SI{5.85}{\dB}. For the \SI{20.5}{\hour} mission orbit, ground-station visibility exhibits an approximately 6-day near-repeat pattern, with access-window durations varying substantially among stations. Most notably, as the encounter geometry places START and Apophis over the Atlantic Ocean, the Cape Verde station provides a continuous visibility window exceeding two days and spanning the flyby, enabling uninterrupted TT\&C coverage throughout the critical encounter phase.}

\subsection{\rev{Satellite management unit}}
% Wang Sun (100 words)

\rev{The SMU serves as the core on-board computing and management unit of the spacecraft. It employs a radiation-hardened SPARC V8 processor, with external synchronous dynamic random-access memory (SDRAM) and static random-access memory (SRAM) used for program execution and NOR flash memory used for boot software storage. The processor accesses a field-programmable gate array (FPGA) through its external memory interface, while the FPGA acts as the central bridge for NAND flash mass-memory mapping, functional expansion, and communication interfaces. During operation, the processor performs reliability maintenance of external storage resources, including periodic maintenance of SDRAM and SRAM, verification of NAND flash data, and monitoring and management of peripheral address spaces. The FPGA further provides watchdog, reset, interrupt, time, and pulse-per-second management functions. This processor--FPGA architecture enhances the robustness and autonomy of the spacecraft's on-board computing and management functions.}

\section{Summary}

The START mission provides a rapid-response, low-cost architecture to investigate the near-Earth asteroid (99942) Apophis during its 2029 Earth close approach. By launching as a secondary payload to low Earth orbit and utilizing a solar electric propulsion system for orbit raising, the \rev{\SI{300}{\kg}} spacecraft will intercept Apophis \rev{near its ECA, when its geocentric distance is approximately \SI{38000}{\km}}. This trajectory enables a high-speed flyby of Apophis at a closest distance of \SI{7}{\km}, timed to coincide with Apophis's ECA.
During the encounter, the payload suite, consisting of narrow- and wide-field cameras alongside \rev{three near-infrared multispectral imagers}, will acquire centimeter-scale imagery and \rev{meter-scale near-infrared spectra}. \rev{These observations are designed to capture the immediate physical and dynamical responses of a rubble-pile asteroid to planetary tidal forces, including localized mass flow and spectral changes.}
Ultimately, START will yield temporally and geometrically complementary data to international flagship missions. Furthermore, the mission will demonstrate the viability of utilizing commercial launch opportunities and miniaturized spacecraft platforms for asteroid reconnaissance, establishing a scalable model for future planetary defense initiatives.

%% Please use the acknowledgment and contribution environments. This will 
%% be anonomyized when the "anonymous" style option is used. 
\begin{acknowledgments}

This work is supported by the Tsinghua University Disruptive Innovation Talent Development Program, the Tsinghua Dushi Funds, and the Xingjian College (College of Future Technology) Future Technology Challenge Project. \rev{BC was supported by the Beijing Young Talent Nurturing Program.}

\rev{We thank the Deep Space Exploration Laboratory and GALAXYSPACE for their collaboration in the development of the START mission and spacecraft.}

\end{acknowledgments}

\begin{contribution}
%%This section gives authors the space to recognize author contributions. The text inside this environment is NOT counted towards the total word quanta. At a minimum, manuscripts are expected to include this text:

YXW and YDL contributed equally, leading mission architecture and manuscript writing. 
YFJ and WYD developed the mission concept and scientific objectives. 
NZ and JLS led the overall system engineering. 
BWZ, JXX, WS, HZD, SYW, ZXZ, and HYL designed the scientific payload and satellite subsystems. 
YHY, RCZH, JTL, YCS, YTS, TCW, JY, ZJT, and YHW conducted mission analyses.
ZKW, TSW and JFL provided overall guidance and support.
BC and HXBY conceived the START mission and oversaw the overall execution of the project team.

%% But authors are expected to provide more specific details, e.g. 
%%
%%SC was responsible for writing and submitting the manuscript.
%%WWM came up with the initial research concept and edited the manuscript.
%%OTS obtained the funding and edited the manuscript.
%%EBF provided the formal analysis and validation. He also edited the manuscript.
%%GEH Supervised the undergraduates, wrote the software and administers the project github and Zenodo repositories.
%%
%% Authors can use the Contributor Role Taxonomy (CRediT) at
%% for ideas on how write a good statement tailored to their needs.

\end{contribution}


% \appendix

% \section{Appendix information}

% Appendices can be broken into separate sections just like in the main text.
% The only difference is that each appendix section is indexed by a letter
% (A, B, C, etc.) instead of a number.  Likewise numbered equations have
% the section letter appended.  Here is an equation as an example.
% \begin{equation}
% I = \frac{1}{1 + d_{1}^{P (1 + d_{2} )}}
% \end{equation}
% Appendix tables and figures should not be numbered like equations. Instead
% they should continue the sequence from the main article body.

%% For this sample we use BibTeX plus aasjournalv7.bst to generate the

\bibliography{sample701}{}
\bibliographystyle{aasjournalv7}



\end{document}